
\input harvmac


\def\VEV{VEV}
\noblackbox
{\parskip=0pt	
\baselineskip=22pt      

\Title{\vbox{\baselineskip12pt{\hbox{CTP-TAMU-73/92}}%
{\hbox{hep-ph/9211215}}}}
{\vbox{\centerline{Vortex Solution in a Multi-Higgs System}\vskip 2pt
\centerline{and Its Physical Implication}}}
\centerline{HoSeong ~La\footnote{$^*$}{%
e-mail address: hsla@phys.tamu.edu, hsla@tamphys.bitnet}   }

\bigskip\centerline{Center for Theoretical Physics}
\centerline{Texas A\&M University}
\centerline{College Station, TX 77843-4242, USA}
\vskip 0.7in

Classical vortex solutions in $(1+2)$-dimensional
multi-Higgs systems are studied.  In particular
the existence of such a solution requires
equal characteristic lengths and a specific relation between
the ratio of the two Higgs vacuum expectation values
and the couplings in the Higgs potential, if $\lambda_3\neq 0$.
} 		
\baselineskip=22pt   

\bigskip

\Date{11/92} 
 \noblackbox

\def\hat{\widehat}

\def\la{\lambda}
\def\half{{\textstyle{1\over 2}}}

\def\e{{\rm e}}
\def\pa{\partial}
\def\mbox#1#2{\vcenter{\hrule \hbox{\vrule height#2in
		\kern#1in \vrule} \hrule}}  

\font\cmss=cmss10 \font\cmsss=cmss10 scaled 833
\def\IZ{\relax\ifmmode\mathchoice
{\hbox{\cmss Z\kern-.4em Z}}{\hbox{\cmss Z\kern-.4em Z}}
{\lower.9pt\hbox{\cmsss Z\kern-.4em Z}}
{\lower1.2pt\hbox{\cmsss Z\kern-.4em Z}}\else{\cmss Z\kern-.4em Z}\fi}
\def\IR{\relax\ifmmode\mathchoice
{\hbox{\cmss I\kern-.5em I}}{\hbox{\cmss R\kern-.5em R}}
{\lower.9pt\hbox{\cmsss I\kern-.5em I}}
{\lower1.2pt\hbox{\cmsss R\kern-.5em R}}\else{\cmss R\kern-.5em R}\fi}

\def\cos{{\rm cos}}
\def\sin{{\rm sin}}

\def\CL{{\cal L}}

\vfill\eject


Recent measurements of the gauge couplings\ref\PDG{Particle Data Book,
Phys. Rev. {\bf D45} (1992).}\ have led to a growing anticipation
that the minimal supersymmetric Grand Unified Theories (GUTs)\ref\rSGUT{S.
Dimopoulos and H. Georgi, Nucl. Phys. {\bf B193} (1981) 150; N. Sakai, Zeit. f.
Phys. {\bf C11} (1981) 153.}\
or supergravity GUTs\ref\rSGGT{A.H. Chamseddine, R. Arnowitt and P. Nath, Phys.
Rev. Lett. {\bf 49} (1982) 970; L.E. Ibanez, Phys. Lett. {\bf 118B} (1982) 73;
J. Ellis, D.V. Nanopoulos and Tamvakis, Phys. Lett. {\bf 121B} (1983) 123;
K. Inoue, A. Kakuto, H. Komatsu and S. Takeshita, Prog. Theo. Phys. {\bf 68}
(1982) 927; L. Alvarez-Gaum\'e, J. Polchinski and M.B. Wise, Nucl. Phys. {\bf
B221} (1983) 495; J. Ellis, J.S. Hagelin, D.V. Nanopoulos and Tamvakis, Phys.
Lett. {\bf 125B} (1983) 275; L. Iba\~nez and C. Lopez, Nucl. Phys. {\bf B233}
(1984) 545; L.E. Iba\~nez, C. Lopez and C. Mu\~nos, Nucl. Phys. {\bf B250}
(1985) 218.}\ref\revSGUT{For reviews see P. Nath, R. Arnowitt and A.H.
Chamseddine, ``{\it Applied N=1 Supergravity}," (World Sci., Singapore,
1984); H.P. Nilles, Phys. Rep. {\bf 110} (1984) 1; H. Haber and G. Kane, Phys.
Rep. {\bf 117} (1985) 75.}\
with the supersymmetry scale of order 1 TeV or below may be a
phenomenologically plausible unified theory of strong and electroweak
interactions\ref\rUNCoup{P. Langacker, in {\it Proc. PASCOS 90 Symposium} ed.
by P. Nath and S. Reucroft (World Sci., Singapore, 1990); P. Langacker and M.
Luo, Phys. Rev. {\bf D44} (1991) 817; J. Ellis, S. Kelley and D.V. Nanopoulos,
Phys. Lett. {\bf 249B} (1990) 441; {\bf 260B} (1991) 131; U. Amaldi, W. de Boer
and H. F\"ursteanu, Phys. Lett. {\bf 260B} (1991) 447.}.

These supersymmetric GUTs in general require at least two Higgs
multiplets for the electroweak symmetry breaking\revSGUT%
\ref\rtwoH{N.G. Deshpande and E. Ma, Phys. Rev. {\bf D18} (1978) 2574;
 J.F. Donoghue and L.-F. Li, Phys. Rev.
{\bf D19} (1979) 945; H.E. Haber, G.L.Kane and T. Sterling, Nucl. Phys. {\bf
B161} (1979) 493; E. Golowich and T.C. Yang, Phys. Lett. {\bf 80B} (1979) 245.
}\ref\GerH{H. Georgi, Hadronic J. {\bf 1} (1978)
155.}\ref\revHig{For a review see J.F. Gunion, H.E. Haber, G. Kane and S.
Dawson, ``{\it The Higgs Hunter's Guide}," (Addison-Wesley, 1990).}. Each Higgs
gets its own vacuum expectation value (\VEV), say $v_1, v_2$,
to spontaneously break the $SU(2)\times U(1)_Y$ symmetry
down to the $U(1)_{em}$. These \VEV s are phenomenologically
important
but unfortunately they are not determined theoretically except in some no-scale
models\ref\rNano{A.B. Lahanas and D.V. Nanopoulos, Phys. Rep. {\bf 145} (1987)
1; Also see, for example, J.~Lopez, D.V.~Nanopoulos and A.~Zichichi,
Texas A\&M preprint, CTP-TAMU-68/92 (1992).}.
The geometric sum $v^2=v_1^2+v_2^2$ can be determined in terms of the
mass of the gauge boson. This however leaves the ratio of the two \VEV s,
$\tan\beta\equiv v_2/v_1$,  still
undetermined. Thus it is very important to look for any condition which
constrains the ratio rather theoretically, if possible.

With such a motivation in mind, in this letter we shall attempt to find any
property that constrains $\tan\beta$ in multi-Higgs system, e.g.
two-Higgs systems. The result is indeed positive
and we find that there is a simple formula to express $\tan\beta$ in terms of
the couplings of the Higgs potential, so far as a certain vacuum defect exists.
To demonstrate how it works we shall work on a simple (1+2)-dimensional toy
model, but the generalization for a realistic model should be straightforward,
which will be presented elsewhere\ref\MYREAL{H.S. La, to be published.}.

The model we consider here is
a local $U(1)$ gauge theory with two Higgs singlet scalars.
The Higgs
potential we use in fact is motivated by the general two-Higgs potential that
induces $SU(2)\times U(1)_Y\to U(1)_{em}$ symmetry breaking, which is usually
written in terms of two Higgs doublets\GerH\revHig.
In terms of singlet Higgs scalars we shall take the following potential:
\eqn\ehigpot{V(\phi_1,\phi_2)={\la_1\over 4}(|\phi_1|^2-v_1^2)^2+
{\la_2\over 4}(|\phi_2|^2-v_2^2)^2+{\la_3\over
4}(|\phi_1|^2+|\phi_2|^2-v^2)^2,}
where $v^2=v_1^2+v_2^2$.
This two-Higgs system has not only the local $U(1)$
symmetry but also accidental
global $U(1)_1\times U(1)_2$ symmetry respectively for
$\phi_1$ and $\phi_2$\ref\Wein{Such a phenomenon that the potential has a
larger global symmetry than the gauge symmetry was called accidental
symmetry first by Weinberg in S. Weinberg, Phys. Rev. Lett. {\bf 29} (1972)
1698; Also see S. Coleman, ``{\it Aspects of Symmetry}" (Cambridge, 1985).
}.

Note that if $\la_1=\la_2=0$, this Higgs system has an accidental $SU(2)$
global symmetry if we require that $(\phi_1, \phi_2)$ form an $SU(2)$
doublet. This pattern of symmetry breaking, $SU(2)_{{\rm global}} \times
U(1)_{{\rm local}} \to U(1)_{{\rm global}}$, is known to lead to semilocal
topological defects\ref\VaAc{T. Vachaspati and A. Ach\'ucarro, Phys. Rev.
{\bf D44} (1991) 3067; T. Vachaspati, Phys. Rev. Lett. {\bf 68} (1992) 1977.}.
If $\la_3=0$, this becomes simply a decoupled two-scalar system with global
$U(1) \times U(1)$ symmetry. There is a vortex solution trivially
generalized from the result in ref.%
\ref\NieOl{H.B. Nielsen and P. Olesen, Nucl. Phys. {\bf B61} (1973) 45;
E.B. Bogomol'nyi, Sov. J.
Nucl. Phys. {\bf 24} (1976) 449; H.J. de Vega and F.A. Schaposnik, Phys. Rev.
{\bf D14} (1976) 1100.}. But the system we consider here is different from
them.
In this letter we shall stick to the general case that $\la_3\neq 0$ and
at least one of $\la_1$ or $\la_2$ is not zero. Then we shall find that this
system reveals a rather interesting result.

The key observation is that the spontaneous symmetry breaking of Eq.\ehigpot\
leads to a vortex solution, whose existence will introduce an extra condition
on the Higgs \VEV s.
Then we can determine them completely, which are related only by $v^2 =
v_1^2 + v_2^2$ otherwise.

Consider the Lagrangian density in (1+2)-dimensional space-time
\eqn\elagr{\CL=-{\textstyle{1\over 4}}F^{\mu\nu}F_{\mu\nu}
+\half|D_\mu\phi_1|^2+\half|D_\mu\phi_2|^2
	-V(\phi_1, \phi_2),}
where $F_{\mu\nu}=\pa_\mu A_\nu-\pa_\nu A_\mu$ and $D_\mu=\pa_\mu-ieA_\mu$.
Then the equations of motion are
\eqna\eomi
$$\eqalignno{D^\mu D_\mu\phi_1 &+(\la_1+\la_3)(|\phi_1|^2-v_1^2)\phi_1
		+\la_3(|\phi_2|^2-v_2^2)\phi_1=0, &\eomi a\cr
	D^\mu D_\mu\phi_2 &+(\la_2+\la_3)(|\phi_2|^2-v_2^2)\phi_2
		+\la_3(|\phi_1|^2-v_1^2)\phi_2=0, &\eomi b\cr
\pa^\mu F_{\mu\nu} &= J_\nu\equiv J_{1\nu} + J_{2\nu}, &\eomi c\cr
J_{i\nu} &=-\half ie (\phi_i^*\pa_\nu\phi_i-\phi_i\pa_\nu \phi_i^*)
-e^2 A_\nu |\phi_i|^2, \ \ i=1,2. \cr}
$$

For time-independent solutions we choose $A_0=0$ gauge, then the system
effectively reduces to a two-dimensional one. In this case since
we are interested in vortex solutions in $\IR^2$,
it is convenient to represent them in the polar coordinates
$(r, \theta)$\NieOl\ such as
\eqn\eans{\phi_1=\e^{im\theta} f(r),\ \  \phi_2=\e^{in\theta} g(r),\ \
	{\vec A}={\hat e_\theta} {1\over r}A(r),}
where $m,n$ are integers identifying each winding sector.
To become  desired finite-energy
defects located at $r=0$ these should satisfy the
following boundary conditions:
\eqn\ebc{\eqalign{&f(0)=0,\ \ g(0)=0,\ \ A(0)=0,\cr
	&f\to v_1,\ g\to v_2, \ A\to {\rm const.}\ \ {\rm as}\ \
r\to\infty.\cr}}
The constant for the asymptotic value of $A$ will be determined properly later.

In the polar coordinates the equations of motion Eq.\eomi{a-c}\
can be rewritten as
\eqna\eomii
$$\eqalignno{
-{1\over r}\pa_r(r\pa_r f) +{1\over r^2}f(m-eA)^2 + (\la_1 +\la_3) (f^2-v_1^2)f
+\la_3 (g^2-v_2^2)f &=0,\ \ \  &\eomii a\cr
-{1\over r}\pa_r(r\pa_r g) +{1\over r^2}g(n-eA)^2 + (\la_2 +\la_3) (g^2-v_2^2)g
+\la_3 (f^2-v_1^2)g &=0,\ \ \  &\eomii b\cr
-\pa_r^2 A
+ {1\over r}\pa_r A-e\left[(m-eA) f^2 + (n-eA) g^2\right] &=0.\ \ \
&\eomii c\cr}
$$
In general it will be a formidable task to solve these equations exactly, but
luckily, for our purpose it turns out to be
 good enough to find approximate solutions for large $r$.
Imposing the boundary conditions at large $r$,  Eqs.\eomii{a,b}\ become
consistent only if $m=n$ and that it fixes the asymptotic value
$ A\to {n\over e}$ as $r\to\infty$.
This implies that there is no vortex solution of different winding numbers for
different Higgs fields. This is in fact an anticipated result because the
vortex solution we are interested in is due to the spontaneous symmetry
breaking of the local $U(1)$.
With this condition of winding numbers we can solve
Eq.\eomii{c} for large $r$ to obtain
\eqn\solA{A\to {n\over e}-n\sqrt{{\pi v\over 2e}}{\sqrt r} \e^{-r/\la}+\cdots,}
where $\la=1/ev$ is the characteristic length of the gauge field.
Note that the characteristic length defines the region over which the field
becomes significantly different from the value at the location of the defect.

Now  let us determine the characteristic lengths for
$\phi_1$ and $\phi_2$ again for large $r$ as follows.
For simplicity we consider $n=1$ case, but the result does not
really depend on $n$.
Asymptotically we look for solutions of the form
\eqn\eI{f \sim v_1(1- c_1\e^{-r/\xi_1}),\ \ g\sim v_2(1-c_2\e^{-r/\xi_2}),}
where the coefficients $c_1$ and $c_2$  are dimensionless constants.
In principle these constants were calculable if exact solutions were known.
However, solving the equations asymptotically does not necessarily
determine them.
The
exact result however would not change the essential part of the following
arguments because they at most will depend on $\la_i$'s only. (The case that
$c_i$'s depend on $\tan\beta$ to nullify my argument will be momentarily ruled
out below.) Then in the leading order we obtain
\eqna\eII
$$\eqalignno{v_1\e^{-r/\xi_1}
\left[-{1\over \xi_1^2}+2(\la_1+\la_3)v_1^2\right] +2 c\la_3 v_1 v_2^2
\e^{-r/\xi_2}+\cdots &=0 &\eII a\cr
c v_2\e^{-r/\xi_2}
\left[-{1\over \xi_2^2}+2(\la_2+\la_3)v_2^2\right] +2\la_3 v_1^2 v_2
\e^{-r/\xi_1}+\cdots &=0 &\eII b\cr}
$$
where $c\equiv c_2/c_1$ and
the ellipses include terms which vanish more rapidly as $r\to\infty$.

If $\xi_1\neq\xi_2$, then $\la_3$ should vanish. Therefore
to have any nontrivial solution for $\la_3\neq 0$ we are forced to identify
$$\xi\equiv\xi_1=\xi_2,$$
and that by demanding the vanishing coefficient
of $\e^{-r/\xi}$ in Eqs.\eII{a,b}\ we obtain
\eqn\esire{{1\over 2\xi^2}=(\la_1+\la_3)v_1^2 +c\la_3 v_2^2= (\la_2+\la_3)
v_2^2 + {1\over c}\la_3 v_1^2.}
Then we can solve for $\tan\beta$ as
\eqn\edrat{\tan\beta\equiv {v_2\over v_1}=\sqrt{{\la_1+\la_3-{1\over c}\la_3
\over \la_2+\la_3-c\la_3}}.}
If $c=1$, then it simplifies to lead to
\eqn\eresult{\tan\beta=\sqrt{{\la_1\over \la_2}},\ \ \  \la_3\neq 0.}

\def\tc{c}
Suppose Eq.\edrat\ were an identity which did not determine $\tan\beta$ but it
merely computed $c$, $c$ should depend on $\tan\beta$. If so, then one can
solve the following
quadratic equation to obtain $c$ in terms of $\la_i$'s and $\tan\beta$:
\eqn\enorii{\la_3 \tc^2 + \left(\tan^2\beta(\la_2+\la_3) -
(\la_1+\la_3)\right)\tc -\la_3\tan^2\beta=0}
and one can obtain
\eqn\enoiii{\tc_\pm\!=\!{1\over 2\la_3}\!\left[-\tan^2\!\beta(\la_2+\la_3)
+ (\la_1+\la_3)
\pm\!\sqrt{\left(\tan^2\!\beta (\la_2+\la_3)- \!(\la_1 +\la_3)\right)^2 \!+\!
4\la_3^2\tan^2\!\beta}\right].}
Now one can easily see that $c$ is ill-defined for at least two reasons,
unless Eq.\eresult\ or so is satisfied.

Frist, as $\la_3\to 0$, $c$ increases indefinitely or approaches to zero.
One may naively say that this simply implies that
$\xi_1$ and $\xi_2$ are not necessarily the same for $\la_3=0$. However, it is
not the case. The reason is as follows:
If $c$ either increases indefinitely or approaches to zero, at least
one of $c_i$ should approach to zero. Otherwise it makes the asymptotic
formular ill-defined.
Then the asymptotic formula should make sense for sufficiently small
$r\ll\xi$ because this term is already small enough.
Thus we in fact lose the notion of the characteristic length.

Secondly, for $\la_3\neq 0$ as $\tan\beta\to 0$ or $\infty$. Note that if one
uses $\tan\beta$ as an input parameter, then $c$ should be well-defined for any
$\tan\beta$. However, in this case we can easily see that $c$ again approaches
to zero or infinity, unless eq.\eresult\ is satisfied.

Thus Eq.\edrat\ is not an equation to determine $c$. We can get a consistent
result only if $c$ is used as an independent input parameter.
Therefore, one can reasonably assume $c=1$ in a good approximation to obtain
Eq.\eresult\ as our result.

Thus we have shown that the existence of vortex requires a specific
ratio of the two Higgs \VEV s in terms of the
couplings in the Higgs potential. This tells us that although different Higgs
field gets different \VEV s, their characteristic lengths should be the same to
form a single defect. Both Higgs should reach the true vacuum at the same
distance. To do that the two \VEV s should satisfy a proper relation, which is
Eq.\eresult.

Furthermore, together with $v$, we can completely determine the \VEV s as
\eqn\eIII{
v_1 =v\,\cos\beta=v{\sqrt{{\la_2\over \la_1+\la_2}}},\ \
v_2 =v\,\sin\beta=v{\sqrt{{\la_1\over \la_1+\la_2}}}.}
The characteristic lengths $\xi_1, \xi_2$ now satisfy
\eqn\echar{\xi\equiv\xi_1=\xi_2={1\over {\sqrt{2}}v}
\sqrt{{\la_1+\la_2\over \la_1\la_2+\la_2\la_3+\la_3\la_1}}.}
Note that, although $\tan\beta$ does not depend on $\la_3$, it is crucial to
have nonvanishing $\la_3$ coupling to obtain such a result.
The gauge boson mass is $M_A=1/\la=ev$
after spontaneous symmetry breaking. In a realistic model this gauge field in
fact may be identified with a massive neutral gauge boson, e.g. $Z^0$.

Without loss of generality we can assume $v_2 \geq v_1$ or $\tan\beta\geq 1$.
If $\la_1=\la_2$, $\tan\beta=1$. In this case two Higgs are indistinguishable
and we may be less motivated to have such a system. But if we keep in mind of
realistic models in which the couplings of the
two Higgs to fermions are different, there is no reason we stick to this case.
{}From the beginning this indeed has been our main motivation.
Thus interesting cases are when
$\la_1>\la_2$ so that $\tan\beta>1$. To have large $\tan\beta$
the quartic coupling of $\phi_1$ should be much stronger than that of $\phi_2$.
It will be interesting to see how this would
affect a Coleman-Weinberg type analysis\ref\rCW{S. Coleman and E. Weinberg,
Phys. Rev. {\bf D7}(1973)1888.}\
of the Higgs potential given in  Eq.\ehigpot.
In the case $\la_1\gg\la_2$ we observe that $v_2$ is almost the same as $v$,
while $v_1$ becomes much smaller.

So far we have not mentioned anything about the stability of the vortex
solution obtained in this model.
In a special case in which the gauge coupling is related to some of the
Higgs couplings, this vortex solution saturates the Bogomol'nyi
bound. Thus it is a stable finite energy solution\ref\mydhig{
For the detail, see H.S. La, Texas A\&M preprint, CTP-TAMU-1/93.}.
Even though it were not stable, it would not not forbid us
from using the argument presented here to constrain
$\tan\beta$ because it does not
depend on the stability of the solution.
What matters is the formation of the defect at some stage.

This completes the proof of the existence of vortex solutions in the two-Higgs
system with the potential Eq.\ehigpot. The consistency condition of such
existence requires that the ratio of the two Higgs \VEV s is constrained by the
Higgs couplings, which otherwise are not related.
Much work is needed to find
exact vortex solutions, but at this moment we still get nontrivial physical
implication with approximate solutions.

Note that such a vortex solution in the $(1+2)$-dimensional space is nothing
but
the cylindrically symmetric
cosmic string solution in the $(1+3)$-dimensional space-time.
Therefore, by amplifying the result given here, one should be able to determine
the phenomenologically important $\tan\beta$ by studying the cosmic string
solution, or perhaps any (topological) defects, related to spontaneous symmetry
breaking of realistic models, e.g. the two-Higgs-doublet standard model or
the supersymmetric GUTs\MYREAL. If such a defect does not exist,there is no
reason why $\tan\beta$ should be such a form. However, if it exists,
$\tan\beta$ would be such a form. The final answer should be left up to the
measurement of $\tan\beta$ experimentally. Furthermore,
since the argument presented here
does not necessarily depend on the stability of the cosmic string, the universe
does not need to have stable cosmic strings to explain
$\tan\beta$ in such a way.


\bigbreak\bigskip\bigskip\centerline{{\bf Acknowledgements}}\nobreak

\par\vskip.3truein

The author would like to thank R. Arnowitt for helpful discussions.
This work was supported in part by NSF grant PHY89-07887 and World Laboratory.


%
\listrefs
\vfill\eject
\bye